\documentclass[aps,superscriptaddress,reprint]{revtex4-2}

\usepackage[utf8]{inputenc}
\usepackage{xcolor}
\usepackage{pst-node}
\usepackage{soul}
\usepackage{graphicx}
\usepackage{amsmath, amssymb, amsthm}
\usepackage{mathtools}
\usepackage{hyperref}
\usepackage[shortlabels]{enumitem}
\usepackage{dsfont}
\usepackage{amsthm}
\usepackage{amssymb}
\usepackage{amstext}
\usepackage{braket}
\usepackage{amsfonts}
\usepackage{nicefrac}
\usepackage{extarrows} 
\usepackage{graphicx}
\usepackage{relsize}
\usepackage{soul}
\usepackage{xfrac} 

\newcommand{\skiptext}[1]{}

\begin{document}

\title{Cost optimized ab initio tensor network state methods:\\industrial perspectives}
 
\author{Andor Menczer}
\affiliation{%
Strongly Correlated Systems Lend\"ulet Research Group,
Wigner Research Centre for Physics, H-1525, Budapest, Hungary
}%
\affiliation{%
Eötvös Loránd University, Budapest, Hungary
}%

\author{\"Ors Legeza}
\email{legeza.ors@wigner.hu}
\affiliation{%
Strongly Correlated Systems Lend\"ulet Research Group,
Wigner Research Centre for Physics, H-1525, Budapest, Hungary
}%
\affiliation{
Institute for Advanced Study,Technical University of Munich, Germany, Lichtenbergstrasse 2a, 85748 Garching, Germany
}
\affiliation{Parmenides Stiftung, Hindenburgstr. 15, 82343, Pöcking, Germany}

\date{\today}

\begin{abstract} 
We introduce efficient solutions to optimize the cost of tree-like tensor network state method calculations
when an expensive GPU-accelerated hardware is utilized.
By supporting a main powerful compute node with additional auxiliary, but much cheaper nodes to store intermediate, precontracted tensor network scratch data, the IO time can be hidden behind the computation almost entirely without increasing memory peak.
Our solution is based on the different bandwidths of the different communication channels, like NVLink, PCIe, InfiniBand and available storage media, which are utilized on different layers of the algorithm.
This simple heterogeneous multiNode solution via asynchronous IO operation has the potential to minimize IO overhead, resulting in maximum performance rate for the main compute unit.
In addition, we introduce an in-house developed massively parallel protocol to serialize and deserialize block sparse matrices and tensors, reducing data communication time tremendously. 
Performance profiles are presented for the spin adapted ab initio density matrix renormalization group method for corresponding
$U(1)$ bond dimension values up to 15400 
on the active compounds of the FeMoco with complete active space (CAS) sizes of up to 113 electrons in 76 orbitals [CAS(113, 76)].
\end{abstract}
\maketitle
\section{Introduction}

Quite recently a parallel implementation of the ab initio Density Matrix Renormalization Group (DMRG) method~\cite{White-1992a},
a subclass of tensor network state (TNS) algorithms~\cite{Schollwock-2005,Schollwock-2011,Szalay-2015a,Verstraete-2008,Orus-2019,Chan-2020,Baiardi-2020,Cirac-2021,Verstraete-2023},
has been introduced
achieving a quarter petaFLOPS performance on a single NVIDIA DGX-H100 GPU node~\cite{Menczer-2024c}.
This incredible computational power has the potential to pave the way for simulating  challenging multi-reference problems in
chemistry~\cite{Swart-2016,Khedkar-2021,Feldt-2022,Pantazis-2009,Sharma-2014,Vinyard-2017,Lubitz-2019,Kerridge-2015,Spivak-2017,Gaggioli-2018,Trond-2011,Pyykk-2012,Reiher-2014a,Tecmer-2016,Wenjian-2020}
or
highly correlated material science~\cite{Zhang-2017}, i.e., to perform large scale high-accuracy ab initio computations routinely on a daily basis for a broad range of disciplines.
On the other hand, massive parallelization of the TNS algorithms on hybrid CPU-GPU architectures is a delicate issue. It requires significantly more complex optimization tasks compared to a CPU only based environment~\cite{Menczer-2023a,Menczer-2023b,Menczer-2024,Xiang-2024}.

The enormous computational power in state-of-the-art GPUs ~\cite{nvidia-a100,nvidia-dgx-h100,gh200,mi300} together with high-speed device-to-device (D2D) communication via NVLink pose new standards for efficient and ideally scalable implementations. 
\begin{figure}
    \centering        \includegraphics[width=0.48\textwidth]{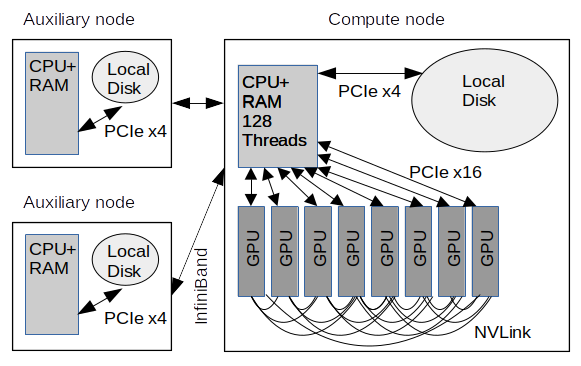}
    \caption{Schematic plot of hardware topology illustrating the various communication channels (arrows), such as host to host (H2H), host to device (H2D), device to host (D2H), and device to device (D2D), i.e., InfiniBand, PCIe, and NVLink, accordingly. The compute node is a very powerful and expensive unit surrounded by one or more cheap auxiliary nodes with minimal computational capacity, but with a substantial amount of RAM or local disk. 
    }
    \label{fig:nodetopology}
\end{figure}
In fact, the different bandwidths associated with the various communication channels, such as  
host to host (H2H), host to device (H2D) and device to host (D2H), and device to device (D2D), i.e., InfiniBand, PCIe, and NVLink (see Fig.~\ref{fig:nodetopology}) must be utilized on different layers of the TNS algorithms.
\begin{figure}
    \centering        \includegraphics[width=0.48\textwidth]{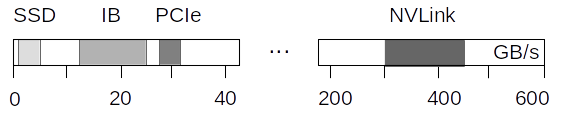}
    \vskip -0.3cm
    \caption{Schematic plot of the different bandwidths associated with different data communication channels measured in GB/s for one-way direction. 
    NVLINK $\simeq 300-450$GB/s (for NVIDIA A100 and H100, respectively), PCIe4.0 x16 $\simeq 32$GB/s,
    InfiniBand $\simeq 12.5-25$GB/s, 
    SSD $\simeq1-5$ GB/s (for write and read operation, respectively).
    Note that NVLink is bidirectional which further doubles the related speed.
    }
    \label{fig:bandwidths}
\end{figure}
In addition, due to differences in available memory sizes in RAM and VRAM the various data sets and computing processes must be carefully organized in order to minimize communication overhead (see Fig.~\ref{fig:bandwidths} for different bandwidths). 
Finally, the various algorithmic parts of the TNS methods utilize different hardware components, thus the total wall time depends on the optimization of each component. 

In the scientific world, the scaling properties of an algorithm in terms of main parameters is the most important question, while in
industrial applications another major factor comes into play: \emph{the cost of the simulations}.
This is especially becoming an important issue nowadays since the price of a utilized hardware
measured in CPU- or in full node-hours,
depending on available memory size, number and type of GPU accelerators, and size of storage media, vary on a large scale in modern high-performance computing (HPC) centers. 
As an example, the price of our previous numerical simulations on the FeMoco and cytochrome P450 (CYP) enzymes~\cite{Menczer-2024c} using a single DGX-H100 node on Google Cloud was around 100 USD per hour.
For ab initio TNS/DMRG calculations with lower accuracy demands, all required data can be kept in memory. Unfortunately, with increasing accuracy threshold, the utilization of additional storage media becomes mandatory in order to save intermediate large scratch data, i.e., precontracted tensor network components on disk. 
In this latter case, the disk IO time can, however, be comparable or even more than the total wall time of all the computation carried out on the GPUs including D2H, H2D and D2D communication time as well. Therefore, optimization of storage IO time or even hiding it behind the computation not only speeds up TNS/DMRG algorithm, but reduces the costs of the simulations drastically.

Distributed memory solutions by employing several compute nodes of the same kind is a possible option, however, the related cost would scale almost linearly with the number of compute units. In contrast to this, in this work, we introduce efficient solutions to optimize the cost of 
ab initio TNS/DMRG calculations, by supporting the main compute node with additional auxiliary, but less powerful nodes to store intermediate precontracted tensor network scratch data.
Therefore, an expensive node with strong GPU support is used to perform the calculations, while intermediate scratch data are transferred via H2H communication to cheaper nodes as part of an asynchronous IO operation. Our solution has the potential to maximize the performance rate of the compute node without increasing the memory peak.

The paper is organized as follows. In Sec.~\ref{sec:method} we present a brief overview of the quantum chemical model Hamiltonian and the DMRG method highlighting those technical aspects which are relevant for efficient data management.
In Sec.~\ref{sec:implementation} we
introduce methods developed according to various parallelization strategies and novel algorithmic solutions to hide data storage time behind computation.
In Sec.~\ref{sec:results} we present numerical benchmarks and scaling analysis for data serialization, synchronous and asynchronous IO operations and
for MPI-based data storage for selected strongly correlated (multi reference)
chemical systems together with discussions on future possibilities.
Point-by-point conclusions, Sec.~\ref{sec:conclusion}, close our presentation.

\section{Method}
\label{sec:method}

\subsection{Numerical procedure}
In the following, our technical solutions will be presented for ab initio tree-like tensor network state methods~\cite{Murg-2010a,Nakatani-2013,Murg-2014,Gunst-2018,Gunst-2019} focusing on the one with the simplest  network topology, namely the density matrix renormalization group (DMRG) method~\cite{White-1992b}. 
Moreover, we focus on a very general form of the Hamiltonian operator 
that can treat any form of non-local interactions related to
two-particle scattering processes. The corresponding
Hamiltonian can be written in the form
\begin{equation}
\mathcal{H} = \sum_{ij\alpha\beta} T_{ij}^{\alpha\beta} 
                c^\dagger_{i\alpha}c_{j\beta} +
                \sum_{ijkl\alpha\beta\gamma\delta} 
      {V_{ijkl}^{\alpha\beta\gamma\delta}
      c^\dagger_{i\alpha}c^\dagger_{j\beta}c_{k\gamma}c_{l\delta}},
\label{eq:ham}
\end{equation}
where the indices
$\alpha,\beta,\gamma,\delta$ label internal degrees of freedom, like spin or isospin.
The operators $c^\dagger_{i\alpha}$ or $c_{i\alpha}$ usually denote spin ladder or
fermion creation and annihilation operators.
Indices $i,j,k,l$ label, in general, arbitrary modes which allows us to simulate strongly correlated quantum many-body problems in various fields of disciplines, like condensed matter physics, nuclear structure theory, or quantum chemistry even in the relativistic domain
~\cite{White-1992b,Xiang-1996,White-1999,Knecht-2014,Dukelsky-2004,Legeza-2015,Legeza-2018a,Shapir-2019,Barcza-2020,Menczer-2024}.

The DMRG algorithm is a variational method in the space of so-called matrix product state (MPS)~\cite{Verstraete-2023} wave functions. In our examples presented for molecular systems, modes are orbitals, and 
an MPS is a parameterization of a many-body wave function in terms of $N$ spinful orbitals $\ket{i_n}$ using $N$ order-3 tensors $A^{i_n}_{\alpha_{n-1}\alpha_n}$ of dimension $(D_{n-1},4,D_n)$, i.e.
\begin{equation}
  \ket{\Psi_{MPS}}  = \sum_{\{i_k\}} \sum_{\{\alpha_p\}}[A_1]_{1\alpha_1}^{i_1} [A_2]_{\alpha_1\alpha_2}^{i_2} \dots [A_{N}]_{\alpha_{N-1}1}^{i_{N}} \ket{i_1\dots i_k}
\label{eq:mps}
\end{equation}
where the first and the last are order-2 tensors or matrices. 
The DMRG method provides a variational approximation to the ground state of Hamiltonian $H$ given by Eq.~\ref{eq:ham} over the space of MPS, i.e. 
\begin{equation}
  E_{opt} = \min_{\ket{\Psi_{MPS}}}\frac{\bra{\Psi_{MPS}}H\ket{\Psi_{MPS}}}{\braket{\Psi_{MPS}|\Psi_{MPS}}}.
\label{eq:opt}
\end{equation}
The ranks of matrices in Eq.~\ref{eq:mps} $D\equiv\max(\{D_n\})$, also called the bond dimension, controls the accuracy of the approximation~\cite{White-1992b,Legeza-2003a,Schollwock-2011,Verstraete-2023,Krumnow-2021} closely related to the truncation error of the DMRG 
~\cite{Legeza-1996,Legeza-2003a} and the Rényi entropy~\cite{Wolf-2008,Eisert-2010}.
In ab initio DMRG calculations, large bond dimension
values of $D\sim\mathcal{O}(10^4)$ are usually mandatory to achieve sufficient accuracy for molecular systems with strong multireference character. The computational complexity and memory requirements of the DMRG with long range interactions scale as $\mathcal{O}(D^3N^4)$  and $\mathcal{O}(D^2N^2)$, respectively. The DMRG algorithm performs an iterative optimization of the wave function by updating MPS tensors one by one solving a large hermitian eigenvalue problem using, e.g. the Lanczos or Davidson method which scales as $\mathcal{O}(D^3)$.
For traditional CPU-based implementations
this step usually accounts for 80\% of the execution time, while in GPU-accelerated versions such ratio can drop to 50\% or less~\cite{Menczer-2023a,Menczer-2023b,Menczer-2024}.
A complete sequence of updates of all tensors is called a DMRG sweep.
During the course of the iterative optimization procedure, precontracted segments of the network must be stored on disk to maximize accessible system sizes and achievable accuracy according to the available RAM size of the compute unit. This, however, can be a major bottleneck for GPU-accelerated simulations as will discussed in the following sections.
For more details on the DMRG and tensor network methods in general, we guide the reader to the existing literature \cite{Schollwock-2005, Schollwock-2011,Noack-2005,Szalay-2015a,Chan-2008,Orus-2014,Baiardi-2020,Verstraete-2008,Cirac-2021}.
\\

\subsection{IO aspects of the DMRG method}

\begin{figure}
    \centering    
\includegraphics[width=0.45\textwidth]{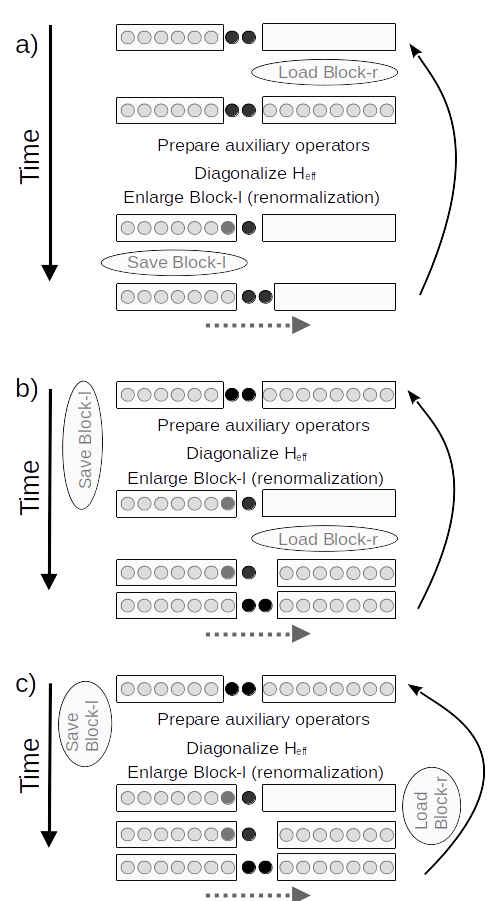}
    \caption{(a) Two-site DMRG topology highlighting precontracted parts of the network, i.e. modes are partitioned into so-called left and right blocks and two modes in the middle which are treated exactly. 
    Scratch data IO procedures for a forward sweep using  synchronous IO implementations. The empty box indicates that the right-block data are cleared from memory.
    The dotted arrow shows the direction of forward sweep.
    (b)In an asynchronous save IO implementation, the save IO operations are performed in parallel to the computational tasks. (c) In an asynchronous IO implementation, both the save and load IO operations are performed in parallel to given set of computational tasks. Arrows with shorter lengths on the left for (a), (b) and (c) indicate systematic reduction of wall time.
    }
    \label{fig:dmrg}
\end{figure}

\begin{figure}
    \centering    
    \includegraphics[width=0.48\textwidth]{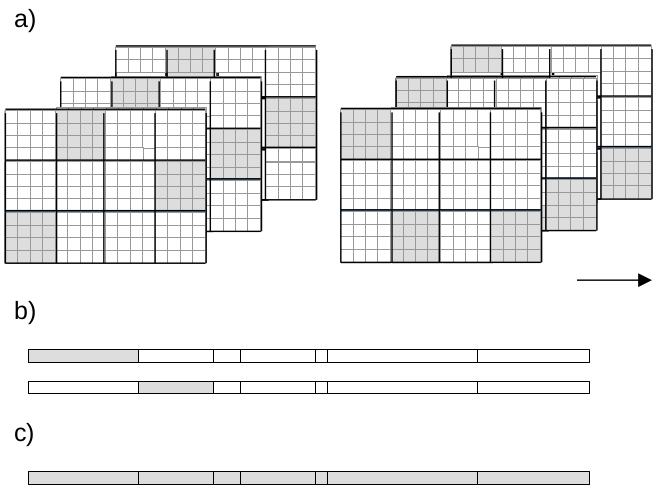}
    \caption{a) Schematic plot of the quantum number-based block sparse representation of matrices and tensors. Shaded area indicates used sectors that are processed in computation and data serialization. Layers indicate position dependent operators which have the same block-sparse structure, but the dense sector matrices take different values for the different layers. 
    b) Skeleton of serialized data segments used during disk save IO procedure or MPI-based communication. Only one segment is filled with data in a given time (shaded region) and transferred immediately to storage media or to another node. This requires only a small additional memory used for buffering.
    c) Skeleton of serialized data segments filled completely with data when asynchronous save IO procedure is utilized. This leads to a substantial increase in peak memory to store redundant data that is saved to storage media in parallel to subsequent computation tasks.
    }  
    \label{fig:serialization}
\end{figure}

In the two-site DMRG algorithm, modes of a network are partitioned into subsystems (blocks), i.e., into precontracted segments of the network, as illustrated in Fig.~\ref{fig:dmrg}(a). The two subsystems in the middle both contain only one mode and are treated exactly.
The underlying algebra to solve the eigenvalue equation given by Eq.~\ref{eq:opt} is performed according to the tensor products of the operators represented on these subsystems~\cite{Schollwock-2005,Noack-2005,Schollwock-2011,Orus-2014,Szalay-2015a}.
In addition, conservation of quantum numbers provides additional constraints, which reduce the computational complexity of the problem and also allow decomposition of full matrices and tensors into smaller components. Therefore, operators are stored in block sparse representation as illustrated in Fig.~\ref{fig:serialization}(a).
From technical point of view, non-zero sector components of matrices and tensors in such a block sparse representation, shaded regions in Fig.~\ref{fig:serialization}(a), go together with series of meta data, which store information about sector indices, related quantum numbers, and sector dimensions among many others~\cite{Menczer-2023a,Menczer-2023b}.

Focusing on the IO aspects of the DMRG method,
a full DMRG iteration step for a given partitioning of modes, also known as superblock configuration, for given sizes of left and right blocks, 
is decomposed into the following main three algorithmic parts~\cite{White-1992a,Xiang-1996,White-1999}: 
1) First, all the data required for the left or right block (environment block) are loaded from a storage media (disk) to memory. Next, additional auxiliary operators are formed and the corresponding effective Hamiltionian given by Eq.~\ref{eq:ham} is constructed. We refer to such a collection of tasks as preprocessing. 
2) Next, the effective Hamiltonian is diagonalized using the L\'anczos or Davidson methods to find a solution of the minimization problem posed by Eq.~\ref{eq:opt}. 
3) This is followed by postprocessing, where various measurements can be performed based on the obtained eigenfunction and a new basis is formed for the enlarged block of the next iteration step via singular value decomposition (SVD).
In a forward sweep, the left block, also considered as the system block~\cite{White-1992a}, is enlarged by one mode, and related operators are mapped to the new basis via the so-called renormalization procedure (also known as network contraction). These block operators are saved on a storage media (see Fig.~\ref{fig:dmrg}(a))
and reloaded at a later stage of the optimization procedure. 
In the subsequent DMRG iteration step, the cut between the two modes treated exactly is shifted by one to the right, and steps 1 to 3 are repeated. This iterative scheme is illustrated in Fig.~\ref{fig:dmrg}(a) for two subsequent DMRG iteration steps in a forward sweep together with the underlying IO operations. The empty right block following the diagonalization step indicates that the corresponding data have been deleted from memory. For the backward sweep, the right block becomes the system block and is enlarged systematically by shifting the cut to the left, thus the IO operations for the left and right blocks are reversed.
During the sweeping procedure, those block data that are not used in later iteration steps are deleted in order to minimize storage space.

In standard CPU-based DMRG implementations and for Hamiltonians with short-range interactions such disk IO operations can be avoided by keeping all data in memory. Moreover, even in case of limited RAM, it would not lead to significant overhead. In contrast to this, for ab initio DMRG, where number of renormalized operators scale with square of the system size,
such IO overhead can be tremendous,
especially for a GPU-accelerated hardware, where computation power can be hundred times higher.
Therefore, burning expensive computation time on such IO operations, which could well exceed wall time of the GPU computation, would make simulations very inefficient budget-wise.

\section{Minimizing IO overhead}
\label{sec:implementation}

As discussed in the previous section,
to maximize the accessible system sizes and bond dimension values based on the size of the RAM,
the renormalized operators of the enlarged block are saved on disk and reloaded at a later iteration step (see Fig.~\ref{fig:dmrg}). In general, the total wall time for saving the block-sparse matrices and tensors is determined by two main algorithmic steps: (1) serialization of the data, i.e., generating long continuous stripes in memory and (2) transferring such serialized data to a given storage medium. The same holds for loading data except that serialization is replaced by deserialization. Here, we discuss three different solutions to optimize such IO overhead.

\subsection{Data management strategies}

1) The simplest workflow is based on the serialization of data in small segments based on the size of a preallocated buffer and saving each segment on a storage media immediately. 
For a schematic plot, see Fig.~\ref{fig:serialization}(b), where shaded regions indicate serialized data segments stored in the buffer. 
This so-called synchronous IO operation usually requires only a small additional memory overhead, but it halts the computation until all data are saved (see Fig.~\ref{fig:dmrg}(b)). Utilization of several CPU threads can, however, speed up serialization reducing wall time significantly, as will be demonstrated in Sec.~\ref{sec:results}.

2) A more efficient implementation is based on asynchronous IO operations, where data are saved on the storage media in parallel to the computation. Here, first all data are serialized and stored in a full array as indicated by the shaded region in
Fig.~\ref{fig:serialization}(c).
Next, a single CPU thread takes care of saving the serialized data while the remaining CPU threads are assigned to proceed with subsequent computational tasks. Therefore, IO time can be hidden behind the computation leading to a drastic reduction in total wall time of the DMRG calculation. This solution, however, requires additional care when a multithreading serializaton protocol is used. First, to avoid the need of multiple copies of the full array, a pointer structure is initialized to determine and store the resulting grid structure of the serialized data based on the block sparse structure of the matrices and tensors. This grid structure is illustrated in Figs.~\ref{fig:serialization} (b) and (c) by breaking up the entire array into smaller components. The related preallocated buffers are filled with actual data only afterward. Although the
asynchronous IO operation leads to a big improvement in performance, it also increases memory peak substantially. In case of DMRG this accounts for some additional 30\% of the memory required to construct the effective Hamiltonian.

3) In order to maximize performance, i.e., by reducing IO overhead, but without increasing memory consumption, 
the asynchronous IO operation can be executed using an auxiliary node as shown in Fig.~\ref{fig:nodetopology}. 
Here, data of the block sparse matrices and tensors are serialized into small segments like in 1), but such segments are transferred to another auxiliary node immediately and stored in the main memory of that auxiliary node. Therefore, this solution again requires only a small additional memory overhead, but it halts the main calculations for a much smaller time period compared to 1), i.e., until data are serialized and transferred via the H2H communication channel. Moreover, this solution can be further extended by using several auxiliary nodes to store precontracted network scratch data and even asynchronous IO operations can be executed on the auxiliary nodes using their local disks (if available) without halting the computation on the main compute node. Since H2H communication via InfiniBand is much faster than saving data on local storage media, such as SSD, the performance of the powerful main node can be maximized without burning expensive computational time on IO overhead.  

\subsection{TNS/DMRG specific implementations}

\subsubsection{Proper balance between high- and low-level languages}

In our TNS/DMRG implementation complex mathematical models and related data structures are encoded via MATLAB~\cite{MATLAB},
a high-level interpreter-based language for modeling complex systems. On the other hand,  serialization of block sparse matrices and tensors, disk IO and MPI-based IO operations, singular value decomposition, renormalization and diagonalization of the effective Hamiltonian are all implemented in native C++ and CUDA. Low-overhead data exchange between the two worlds is made possible by special pointer structures via MATLAB’s MEX interface.
Therefore, complex structures, but lightweight meta-data, usually a few MB in size, are saved via MATLAB which leads to marginal IO overhead, or it can even be kept in memory. On the other hand, heavy objects like block sparse matrices and tensors occupying several hundreds of GBs or even more, are handled via our in-house developed kernels.

\subsubsection{High performance parallel serialization}

Regarding serialization, initially we developed an implementation relying on C++ Boost~\cite{boostlib} and OpenMP. However, due to poor performance, the former has been replaced with our own serialization library. Unlike many such libraries, our version uses no internal buffers, no handles or auxiliary objects and creates no intermediate data. Using templates and compile time code evaluation the compiler is expected to generate class specific serialization functions, each with no internal function calls or any interaction with objects other than the subject of the serialization. The resulting string is written directly into the output buffer, which, in case of multi node scenarios, also happen to be the input buffer for the MPI operations, thus keeping the flow of data as straightforward and minimalistic as possible.

One of the highlights of our implementation is the low cost and exact prediction of post-serialization data size before the actual serialization takes place. This enables us not only mere buffer preallocations, but gives us the ability to parallelize the serialization of variable sized classes without using separate buffers or relying on intra-operative approaches such as inter-thread communication, shared objects and guarded critical sections, all of which would essentially let the threads slow down, talk to each other and figure out indexing during the course of serialization. We grant our library no such time consuming luxury and instead let the main thread map the output buffer based on the pre-operative size information given by the serialization library. By the time each thread familiarises itself with its new given task, the exact memory interval for each thread within the single output buffer
is known to both the main thread and the corresponding serialization thread.

Conclusively, all threads serialize in parallel and in complete independence. Yet, together they produce a single, continuos stream of bytes, which deserializes into an object identical to the one used as the original input.

\subsubsection{Massively parallel MPI communication}

\begin{figure}
    \centering    
    \includegraphics[width=0.48\textwidth]{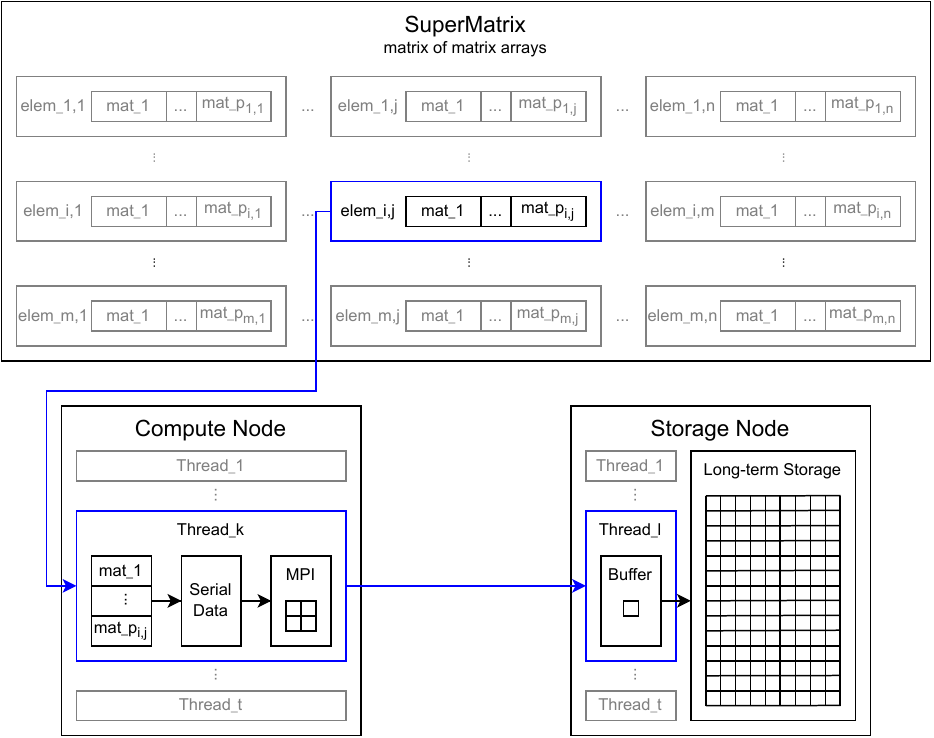}
    \caption{In a multi-node configuration elements of nested arrays are distributed among threads, serialized and optionally further cut down to fit inside reasonably sized MPI packages.}
    \label{fig:mpi_threads}
\end{figure}

When our aim is to prepare, send and receive data between computer nodes, parallel serialization becomes somewhat of a different problem. During serialization the thread is not participating in multi-node logistics as its current operation has no effect on other computers. It is only during the MPI send and receive operations the thread actually moves data between nodes. Because of this, it is pivotal to reach a state in which at least part of the object is serialized as early as possible. 

To conform with the above we once again take advantage of the fact that we know the exact serialization sizes of different parts of complex objects and distribute the components among threads accordingly. Each thread then serializes the received chunk into a temporary buffer and broadcasts the string to the outside world immediately. Soon after the thread requests another, not yet processed chunk of the same object. Processing objects in arbitrarily sized parts decreases latency as well as memory footprint, as the entire byte string never has to make an appearance on the sending side. The schematic plot of the parallel serialization combined with MPI communication is presented in Fig.~\ref{fig:mpi_threads}.

Even if concurrent send operations are not supported by a particular MPI implementation, our parallel model enables threads to concurrently and near instantaneously serialize some parts of the object, thus at least one of the threads can start its MPI operations very early. In the background, all the other threads will continue to chew threw the remainder of the object and be ready to send their own packages once the current sender is finished. This ensures continuous send and receive operations with no idle time.

Each serialized component is self contained in the sense that the string contains not only the deserializable data, but also metadata which pinpoints the location of the component within the original, whole object. As a result, these packages can be sent, mixed with other packages and received back without the need to keep track of their order or the threads they were handled by previously.

Optionally, these packages can be cut down into even smaller packages (see the MPI subprocess in Fig.~\ref{fig:mpi_threads}). This further division serves no other purpose than to keep MPI package sizes below an implementation or self imposed upper limit. The packages produced by the same serialization call are kept together in order to ensure the serialization remains reversible. On the receiving end single packages might be temporarily buffered due to performance reasons, however in the end all packages are stored in a way that the whole, deserializable byte strings are stored continuously in memory.

\subsubsection{Required modifications in case of asynchronous IO operations}

In case of asynchronous IO the order of the load and save operations has to be reorganized in the DMRG method.
In addition, we have to distinguish the different aspects of the asynchronous load and save IO operations. 

First, we discuss the asynchronous save IO procedure as shown in Fig.~\ref{fig:dmrg}(b). In the modified DMRG layout the environment block data of the subsequent DMRG iteration step are loaded already after the renormalization step of the current DMRG step, i.e., before the asynchronous save IO operation is started. This does not increase the memory consumption since right after the diagonalization of the effective Hamiltonian the environment block data are deleted (see the empty right block in Fig.~\ref{fig:dmrg}(b). Therefore, we simply replace the current environment block with the environment block of the subsequent iteration step. After such step, the asynchronous save IO operation can be started.
As long as the wall time of the subsequent DMRG preprocess,  diagonalization and postprocess time together    
is longer than the wall time of saving the renormalized operators, the IO time can be hidden behind the computation. 
In the opposite case, a fetching~\cite{footnote-fetching} is performed before the block data is loaded, thus the compute time can be hidden behind the IO time.

Similarly to the asynchronous save IO operation, the asynchronous load IO operation can also be utilized as summarized in Fig.~\ref{fig:dmrg}(c). If the preprocessing and diagonalization time together is equal to or less than the save IO time, then the environment block data can be loaded via asynchronous IO in parallel of the renormalization step. This means that the block load IO time can also be hidden behind the computation. In this case, however, once the serialized data is loaded a multithreaded deserialization procedure is performed to restore the block sparse matrix and tensor structures.
In general, such deserialization is performed only after the renormalization step is completed in order to provide both algorithmic tasks full access to all threads. Alternatively, available threads can be partitioned between the two tasks.
Here, when a local disk is used a similar increase in peak memory is observed as both serialized and deserialized data appear in the memory during the deserialization step.

In general, the asynchronous load could be executed immediately after the asynchronous save is completed, but that could lead to further increase in peak memory if the diagonalization of the effective Hamiltionian is still running. In this case, data of the left and right blocks of the current DMRG iteration step together with the serialized system block data plus the buffer for the deserialized environment block data of the subsequent DMRG iteration step would all be in the memory. See the asynchronous save and load operations in Fig.~\ref{fig:dmrg}(c) both having their own memory buffer.
If such a high memory peak can be accepted due to large RAM size, then organizing asynchronous load and save into a kind of queuing system together with checkpoints to make sure that required data of a given computational task are already available in the memory, could maximize performance rate regarding IO operations.  

\subsubsection{Required modifications in case of auxiliary nodes}

When an auxiliary node is used to store precontracted tensor network components, the only modification to the above discussed asynchronous save and load operations is that serialized data are not saved to disk, but transferred to the auxiliary node via H2H communication. Similarly, serialized data stored on the auxiliary node can be retrieved again via H2H data transfer when needed and deserialized on the main compute node. Therefore, no computational heavy tasks are executed on the auxiliary node since both serialization and deserialization are performed on the main compute node. As a consequence, cheap and less powerful CPU-based node can be employed to handle IO operations related to the precontracted tensor network components.

For further optimization of storage space, once serialized data are transferred to the auxiliary node the asynchronous save operation can also be executed on the auxiliary node using its local storage media. Similarly, asynchronous load operation can be executed to retrieve data stored on the disk of the auxiliary node and transfer such serialized data to the main compute node. 
See Fig.~\ref{fig:nodetopology} illustrating an auxiliary node with local disk.
In addition, several auxiliary nodes can be used by proper labeling of the stored data sets. The simplest solution in DMRG is to use two auxiliary nodes to store precontracted network components of the left and right blocks.
In our benchmark calculations presented in the following sections we have used up to ten auxiliary nodes.
Here we remark, that the same multithread-based serialization protocol and MPI-based IO transfer are used in our multiNode-multiGPU DMRG variant~\cite{Menczer-2023c}, where the auxiliary nodes are used not only to store precontracted tensor network scratch data, but to perform  computational tasks as well. In that case data received by the auxiliary nodes are deserialized first, but here the auxiliary nodes are also powerful compute units, thus deserialization is also very fast utilizing a large number of threads.
Our solution can be extended easily to more general TNS methods, using more advanced scheduling for loading and saving precontarcted tensor network components.

\section{Numerical results}
\label{sec:results}

In this section, we demonstrate the efficiency of our IO protocols to reduce or even eliminate entirely the IO time related to precontracted tensor network components. Numerical results will be presented for a large chemical complex, namely the FeMoco, as it also serves as the basis for various benchmark calculations~\cite{Reiher-2017,Li-2019,Kai-2020,Brabec-2021,Menczer-2023a,Menczer-2023b,Xiang-2024,Menczer-2023c} due to its important role in nitrogen fixation (i.e., reduction of nitrogen (N$_2$) to ammonia (NH$_3$))~\cite{Hoffman-2014}, which is essential for the biosynthesis of nucleotides such as DNA underlying all life forms on Earth.

\subsubsection{Hardware specifications}

Calculations on CAS(54,54) and CAS(113,76) model spaces have been performed on HPE ProLiant XL675d Server (Hawk-AI) at the High Performance Computing Center Stuttgart (HLRS) and at the Wigner Scientific Computing Laboratory Budapest (WSCLAB). The technical specification of the former one follows as:
Gen4 NVMe disk Hewlett-Packard Enterprise SSD 1.92TB 
(Read speed: 5.37 GB/s 155000 IOPS,
Write speed: 1.12 GB/s 30000 IOPS),
PCIe4 x16 with max theoretical bandwidth 32 GB/s (actual benchmark measurements about 25 GB/s) and
2nd Gen NVLink where the entire NVLink fabric can deliver 600 GB/s, but actual benchmark measurements are about 360GB/s. The server contains two of such SSD disks in a raid-0 configuration thus related performance is doubled. The H2H interconnect deploys an
InfiniBand HDR based interconnect with an 8-dimensional hypercube topology~\cite{hlrs} together with a bandwidth of 20.5 GB/s and an MPI latency of $\sim$1.3$\mu$s per link.
The compute power of the node is based on AMD EPYC 7702 CPUs with $2\times64$ cores supplied with eight NVIDIA A100-SXM4-80GB  devices. The single server at WSCLAB has very similar properties.

\subsubsection{Numerical analysis on CAS(54,54)}

\begin{figure}
    \centering    
    \includegraphics[width=0.48\textwidth]{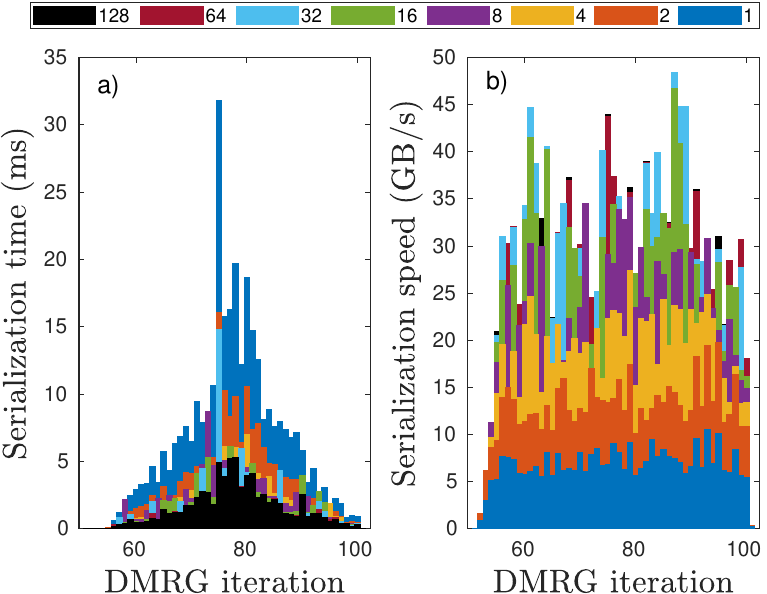}
    \caption{(a) Serialization time in ms
    and (b) serialization performance in GB/s 
    using various number of CPU threads as a function of the DMRG iteration steps in a full DMRG sweep for the FeMoco CAS(54,54) model system using $D=5120$ $SU(2)$ multiplets.
    }
    \label{fig:ser_speed}
\end{figure}

In Fig.~\ref{fig:ser_speed} we show the scaling of the serialization time (a) and the corresponding 
performance (b) measured in seconds and 
GB/s, respectively, as a function of
DMRG iteration steps for a full DMRG sweep using $D=5120$ $SU(2)$ multiplets for the FeMoco CAS(54,54) model space. The tremendous reduction in wall time with an increasing number of threads is obvious. Although for the analyzed data set a saturation in performance is observed for 32 threads, for larger $D$ values and for larger system sizes, this is expected to shift to larger thread numbers.

\begin{figure}
    \centering    
\includegraphics[width=0.48\textwidth]{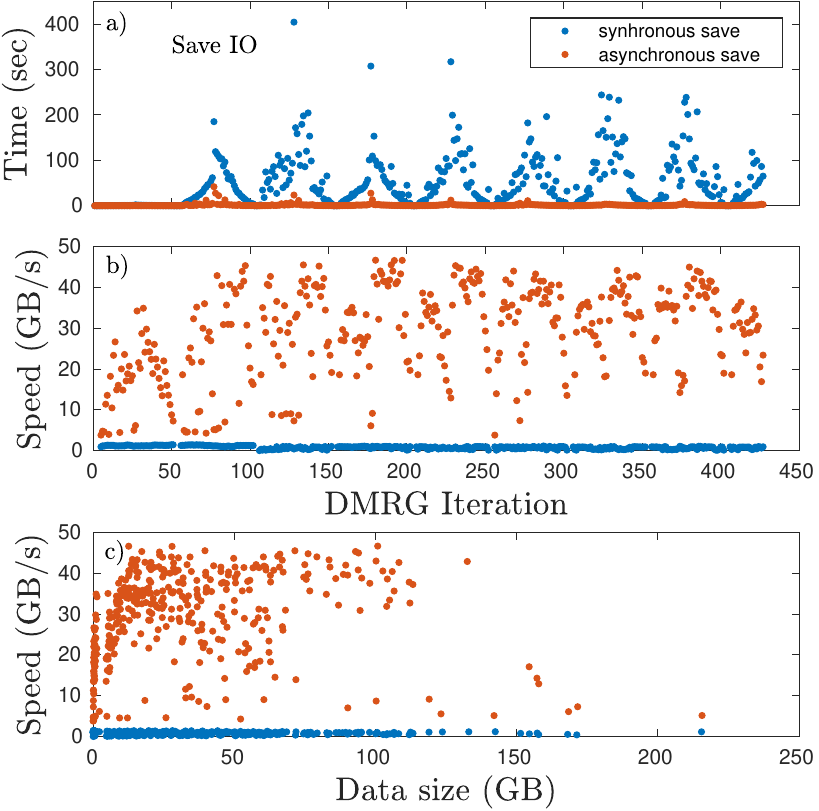}
    \caption{(a) Time and (b) speed of the synchronous and asynchronous save IO operation in sec and
    in GB/s, respectively, as a function of DMRG iteration steps    
    for the FeMoco CAS(54,54) model space with $D=5120$ $SU(2)$ multiplets. 
    (c) Speed plotted as a function of Data size measured in GB.
    The save IO time also
    includes serialization time obtained  via 128 CPU threads.    
    }
\label{fig:save}
\end{figure}

In Fig.~\ref{fig:save} a similar plot is shown for the synchronous and asynchronous save IO procedure including wall time of the serialization step using 128 threads. 
While the wall time for synchronous save using the local disk of the compute node fluctuates between 50 and 400 seconds per DMRG iteration step, it drops below three seconds via the asynchronous save operation (except for a few scattered data points). 
In the latter case, the performance is determined mainly by the serialization time
as is apparent in Fig.~\ref{fig:save}(b), i.e. no fetching is observed. This means that the compute time exceeded the wall time of the save IO time, resulting in a speedup by a factor of 100. Similar performance gains are achieved for the asynchronous load IO operation. As a consequence, the IO time can be hidden entirely behind the compute time with the cost of an increase in the memory peak by some 40-50\%.

Here we remark that efficient serialization and deserialization of the block sparse matrix and tensor components are crucial not to slow down the IO operations.
For example, using MATLAB, the performance of the save IO procedure saturated around 0.53 GB/s while even using only our synchronous save operation we already measured a speed of 2.15 GB/s. 
We also observed stable performance for a broad range of serialized data sizes as illustrated in Fig.~\ref{fig:save}(c).

\begin{figure}
    \centering    
\includegraphics[width=0.48\textwidth]{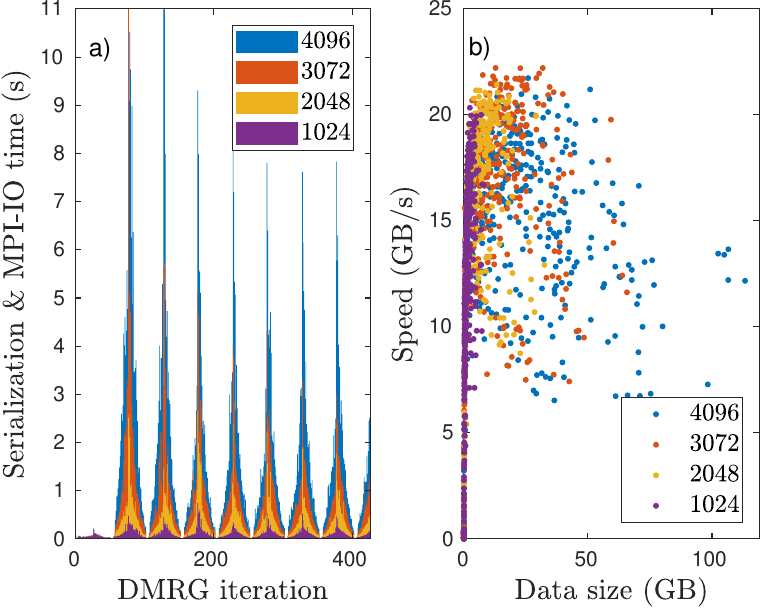}
    \caption{(a) Data transfer time in second including serialization and MPI-IO steps using an auxiliary node as a function of DMRG iteration steps for the FeMoco CAS(54,54)
    and for various $SU(2)$ multiplet $D$ values.
    (b) Related performance in
    in GB/s as a function of the transferred data size measured in GB. The massively parallel serialization has been performed using 128 CPU threads while the available maximum bandwidth of the InfiniBand HDR is 25 GB/s.
    }
    \label{fig:mpisave}
\end{figure}

Next, we turn to our new IO protocol when the compute node is supported by an auxiliary node to store precontracted network components. Here, the related data are transferred between the two nodes via a 25 GB/s InfiniBand, setting also an upper bound on the IO performance. In Fig.~\ref{fig:mpisave}(a)
we present the data transfer time in seconds including serialization and MPI-based IO time 
as a function of DMRG iteration steps for the FeMoco CAS(54,54) model space using various $D=1024,2048,3072,4096$ $SU(2)$ multiplet values.
To provide further insights,
the related performance in GB/s as a function of the transferred data size measured in GB is shown in Fig.~\ref{fig:mpisave}(b).
The massively parallel serialization of the block sparse matrices and tensors has been performed again using 128 CPU threads. 
Although the serialization time increases with $D$ it stays again below a few seconds and a sustainable performance in the range between 14 and 20 GB/s has been realized. Therefore, our heterogeneous multiNode protocol is only slightly slower than the asynchronous IO operation via local disk, but it does not lead to an increase in peak memory. 
This is a major improvement with respect to asynchronous IO operations using local storage (disk) media, since all available RAM can be used to process the effective Hamiltonian maximizing achievable system sizes and bond dimension values on a single compute node. Here we remark, that buffer sizes and number of MPI data packages have a great influence on the IO speed, thus we expect that further optimizations can result in even higher data transfer rate also for large data sets (see the scattered data points for data sizes larger than 40 GB). Again, efficient serialization is a very important issue to suppress related wall time, which might again be a problem in MATLAB Parallel Server which can utilize only up to some 20-30\% of the available H2H bandwidth.

\begin{figure}
    \centering    
\includegraphics[width=0.48\textwidth]{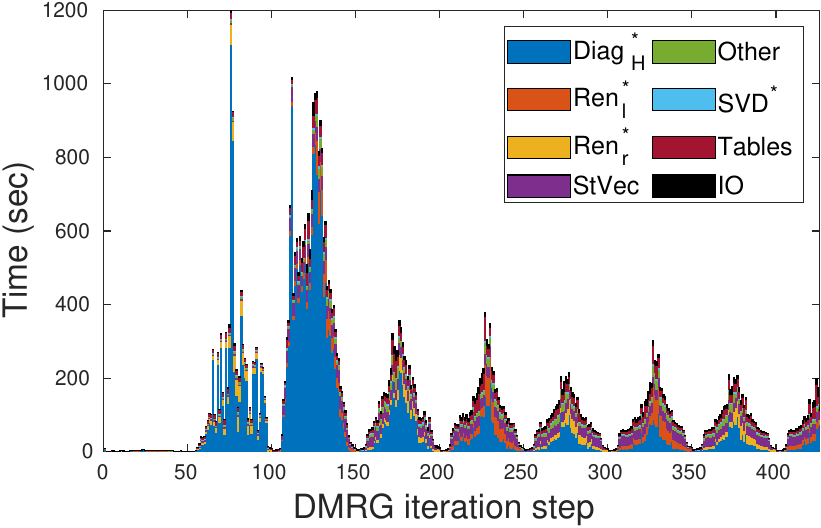}
    \caption{Decomposition of the total wall time 
    as a function of DMRG  iteration steps
    for the FeMoco CAS(54,54) model space using auxiliary nodes and $D=5120$ $SU(2)$ multiplets corresponding to largest $U(1)$ bond dimension values around 17500.
    The asterisks indicate functions converted to GPU already. The description of the legend is given in the main text.
    }
    \label{fig:dmrgtime}
\end{figure}

For completeness, in Fig.~\ref{fig:dmrgtime} we show the decomposition of the total wall time as a function of the DMRG iteration steps using  auxiliary nodes to store the precontracted network components for the FeMoco CAS(54,54) model space using $D=5120$ $SU(2)$ multiplets.
Here, Diag$_{\rm H}$ stands for the diagonalization of the effective Hamiltonian,
Ren$_{\rm l}$ and Ren$_{\rm r}$ for the renormalization of the left and right blocks,
StVec for wave function transformation, i.e., for generation of the starting vector for the diagonalization step, 
SVD for
singular vale decomposition,
Tables for generating meta data for scheduling algorithms, and
IO for all read and write operations, respectively.
Functions already converted to our hybrid CPU-GPU kernel are indicated by asterisks.
It is clearly visible that 
the diagonalization time dominates the total wall time until convergence is reached.
As DMRG converges the number of L\'anczos or Davidson iterations also decreases, thus the wall time of the diagonalization procedure shrinks accordingly.    
In contrast to this, the IO time using the auxiliary node to store precontracted network components is almost invisible for the entire sweeping procedure. 
Furthermore, we expect that the time contribution related to the generation of the starting vector will also decrease by more than an order of magnitude once it is also converted to GPU, similarly to what we have achieved for singular value decomposition.
Contributions of all measurements and non-optimized monitoring functions are collected to ``Other" which can be also accelerated by switching on and off the various individual terms.
Therefore, converting all major algorithmic parts to GPUs and utilizing the various data communication channels on different layers of the algorithm, the DMRG has the potential to scale ideal on high performance computing infrastructures.

\subsubsection{Ultimate need for MPI-based IO operations}

\begin{figure}
    \centering    
\includegraphics[width=0.48\textwidth]{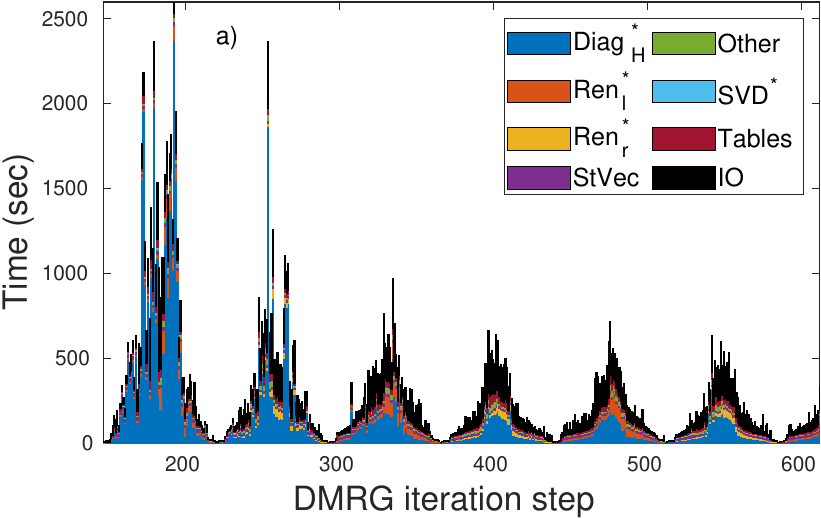}
\includegraphics[width=0.48\textwidth]{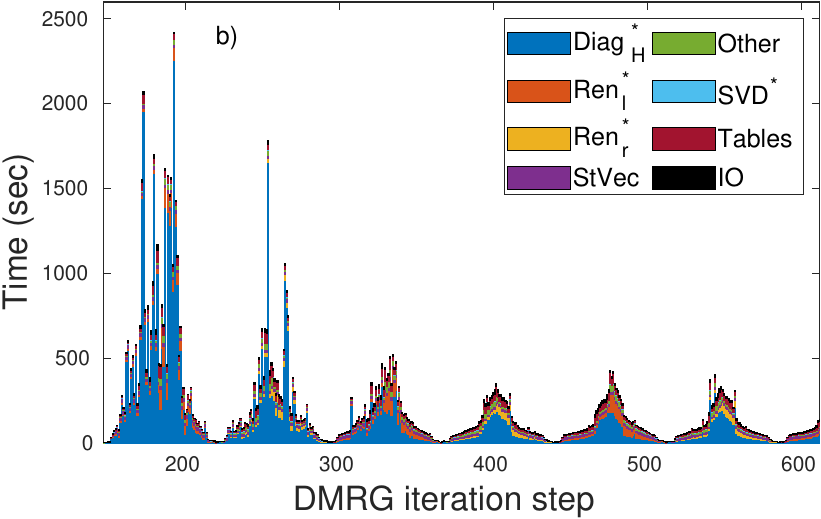}

    \caption{Decomposition of the total wall time 
    as a function of DMRG iteration steps
    via synchronous IO operations (a) and via auxiliary nodes (b)
    for the FeMoco CAS(113,76) model space using $D=4096$ $SU(2)$ multiplets corresponding to largest $U(1)$ bond dimension values around 15400.
    The asterisks indicate functions converted to GPU already. The description of the legend is given in the main text. The first (warmup) sweep with $D=512$ low bond dimension is not shown in the the plots.
    }
    \label{fig:N76_dmrgtime}
\end{figure}

To demonstrate the desperate need for our developments in case of problems with increasing computational complexity, we present the decomposition of the total wall time as a function of DMRG iteration steps 
using synchronous IO operations vial local SSD storage media and via auxiliary nodes in Fig.~\ref{fig:N76_dmrgtime}
for the largest model space considered so far in the literature, i.e. for the FeMoco CAS(113,76) model space~\cite{Li-2019}. 
Here we have used $D=4096$ $SU(2)$ multiplets corresponding to largest $U(1)$ bond dimension values around 15400.
As can be seen in  Fig.~\ref{fig:N76_dmrgtime}(a) when synchronous IO operations are used the IO time dominates the calculation as DMRG converges, which is due to the tremendous speedup of the computation part via GPU accelerators. 
When asynchronous IO is applied (not shown) the IO time has been reduced, but could not be hidden behind the computation for several DMRG iteration steps. This is due to the very large data sets related to precontracted network segments. This can be seen in Fig.~\ref{fig:N76_dmrgtime}(a) for those DMRG iteration steps for which IO time (black) is larger than the sum of all the remaining time components (indicated by different colors). Therefore, a fetching appeared several times before subsequent load and save operations could have been executed, halting the computation for long time periods.
In contrast to this, the IO time almost disappears when auxiliary nodes are used via MPI-based IO operation as demonstrated in Fig.~\ref{fig:N76_dmrgtime}(b).  
Therefore, for larger system sizes and larger bond dimension values, it is mandatory to use auxiliary nodes since the IO time to handle percontracted network components in a given DMRG iteration step exceeds the corresponding compute time significantly. 

\subsubsection{Natural extensions and ideal scalability}

A natural extension of the single auxiliary node version is to utilize two auxiliary nodes for the left and right precontracted network components, respectively. In addition, the number of auxiliary nodes can easily be increased via proper labeling of the transferred data sets. In our pilot implementation, the ID of the compute node was set to zero, while the IDs of the auxiliary nodes to store precontracted network segments of the left and right blocks were determined as
$\rm{mod}(l-1, n/2)+1$ and $\rm{mod}(r-1, n/2)+n/2+1$, respectively, where $l$ and $r$ stands for the size of the left and right block, and $n$ for the number of auxiliary nodes, that we have chosen to be even. 
Since block data sizes increase toward the middle of the DMRG topology, this solution distributes data among the auxiliary nodes uniformly.
In our benchmark calculations, we have used up to ten auxiliary nodes, each with 1TB of RAM.
More sophisticated solutions to distribute data by monitoring available (free) RAM of the auxiliary nodes can also be applied to optimize data storage.  
Moreover, asynchronous read and write IO operations can be performed on the auxiliary nodes to store  precontracted network scratch data on their local disks.

Our solution is not limited to a single compute node, but several nodes with high computational power can be connected. Here, each node is supplied with auxiliary nodes to handle precontracted network segments. In this case, the overall data appearing in the calculations are also divided into smaller segments depending on the available RAM size and distributed among the compute nodes. Therefore, even having a lower upper limit on communication speed due to the finite bandwidth of H2H connection, each compute node deals only with a limited data set.  
Such multiNode-multiGPU asynchronous IO optimized version can raise the performance of DMRG well into the petaFLOPS regime~\cite{Menczer-2023c}.
Moreover, we note that the IO step could be avoided entirely if precontraction of the network segments for a given block size would be executed on-the-fly via sequence of renormalization steps using the stored MPS matrices.
Currently, we have found such solution to be slower than the measured performance via auxiliary nodes, but in the future distributed memory solution together with fast NVLink could lead to major breakthrough regarding this issue as well.

\section{Conclusion}
\label{sec:conclusion}

In this work, we have introduced novel algorithmic solutions together with implementation details to exploit
the various data communication channels, like NVLink, PCIe, InfiniBand and available storage media on different layers of the ab initio density matrix renormalization group method. As demonstrated for the FeMoco CAS(54,54) and CAS(113,76) model spaces ~\cite{Reiher-2017,Li-2019}, 
our simple heterogeneous multiNode solution via asynchronous IO operation has the potential to minimize IO overhead, resulting in a maximum performance rate for the main compute unit. 

We also introduced an in-house developed massively parallel protocol to serialize and deserialize block sparse matrices and tensors speeding up the procedure tremendously.
Moreover, we have discussed in detail the implementation of asynchronous IO operations to store and retrieve precontracted tensor network components using local disk of the compute node. By hiding the IO time behind the compute time entirely, we reported at least a factor of two to three reduction in wall time. This we achieved by paying an extra cost of an increase in memory peak by some 30-50\%.

By replacing the local disk with additional auxiliary nodes in the IO protocol we have presented an efficient solution to handle precontracted network data via MPI-based data transfer. Although this sets an upper bound on the performance based on the technical parameters of the host-to-host (H2H) data communication channel, it does not increase memory peak in the compute node and only slightly slower the asynchronous IO via a local storage media. 
As a consequence, not only the performance can be maximized, but the related simulation costs can be reduced drastically by supporting an expensive GPU-accelerated hardware with much cheaper auxiliary nodes. 
This can be realized via flexible resource allocations, advanced scheduling and queuing options accessible in most of the cloud services.

For completeness, we have presented the decomposition of the total wall time in a full DMRG calculation for the largest systems considered so far in the literature, i.e., for FeMoco CAS(54,54) and CAS(113,76) model spaces.  
Currently in our DMRG algorithm the hybrid CPU multiGPU implementation is available for the diagonalization of the effective Hamiltonian, renormalization of the block operators, singular value decomposition and calculation
of expectation values of operators including one- and two-body reduced matrices. The conversion of other parts is under systematic development, leading to an almost ideal scaling for the DMRG algorithm on high performance computing infrastructures.

The presented framework can easily be extended  to more general TNS methods by using more advanced scheduling protocols to handle IO operations related to preconcerted tensor network components. Our massively parallel IO kernel also serves as the basis of our multiNode-multiGPU DMRG method, raising its performance into the petaflops regime~\cite{Menczer-2023c}. Details of the implementations will be part of a subsequent publication.

\section*{Acknowledgement}
Authors thank for useful comments and discussions with Jeff Hammond, Christian Simmendinger, Mikl\'os Werner and Korn\'el Kap\'as. This research was supported 
by the Hungarian National Research, Development and Innovation Office (NKFIH) through Grant Nos.~K134983 and TKP2021-NVA-04
by the Quantum Information National Laboratory of Hungary, 
and by the Hans Fischer Senior Fellowship programme funded by the Technical University
of Munich – Institute for Advanced Study. 
\"O.L. has also been supported
by the Center for Scalable and Predictive methods
for Excitation and Correlated phenomena (SPEC),
funded as part of the Computational Chemical Sciences Program FWP 70942, by the U.S. Department of Energy
(DOE), Office of Science, Office of Basic Energy Sciences, Division of Chemical Sciences, Geosciences, and Biosciences at Pacific Northwest National Laboratory.
We thank computational support from the Wigner Scientific Computing Laboratory (WSCLAB) and the national supercomputer HPE Apollo Hawk at the High Performance Computing Center Stuttgart (HLRS) under the grant number MPTNS/44246.


\bibliographystyle{achemso}
\bibliography{main}{}

\providecommand{\latin}[1]{#1}
\makeatletter
\providecommand{\doi}
  {\begingroup\let\do\@makeother\dospecials
  \catcode`\{=1 \catcode`\}=2 \doi@aux}
\providecommand{\doi@aux}[1]{\endgroup\texttt{#1}}
\makeatother
\providecommand*\mcitethebibliography{\thebibliography}
\csname @ifundefined\endcsname{endmcitethebibliography}
  {\let\endmcitethebibliography\endthebibliography}{}
\begin{mcitethebibliography}{67}
\providecommand*\natexlab[1]{#1}
\providecommand*\mciteSetBstSublistMode[1]{}
\providecommand*\mciteSetBstMaxWidthForm[2]{}
\providecommand*\mciteBstWouldAddEndPuncttrue
  {\def\EndOfBibitem{\unskip.}}
\providecommand*\mciteBstWouldAddEndPunctfalse
  {\let\EndOfBibitem\relax}
\providecommand*\mciteSetBstMidEndSepPunct[3]{}
\providecommand*\mciteSetBstSublistLabelBeginEnd[3]{}
\providecommand*\EndOfBibitem{}
\mciteSetBstSublistMode{f}
\mciteSetBstMaxWidthForm{subitem}{(\alph{mcitesubitemcount})}
\mciteSetBstSublistLabelBeginEnd
  {\mcitemaxwidthsubitemform\space}
  {\relax}
  {\relax}

\bibitem[White and Noack(1992)White, and Noack]{White-1992a}
White,~S.~R.; Noack,~R.~M. Real-space quantum renormalization groups.
  \emph{Phys. Rev. Lett.} \textbf{1992}, \emph{68}, 3487--3490\relax
\mciteBstWouldAddEndPuncttrue
\mciteSetBstMidEndSepPunct{\mcitedefaultmidpunct}
{\mcitedefaultendpunct}{\mcitedefaultseppunct}\relax
\EndOfBibitem
\bibitem[Schollw\"ock(2005)]{Schollwock-2005}
Schollw\"ock,~U. The density-matrix renormalization group. \emph{Rev. Mod.
  Phys.} \textbf{2005}, \emph{77}, 259--315\relax
\mciteBstWouldAddEndPuncttrue
\mciteSetBstMidEndSepPunct{\mcitedefaultmidpunct}
{\mcitedefaultendpunct}{\mcitedefaultseppunct}\relax
\EndOfBibitem
\bibitem[Schollw\"ock(2011)]{Schollwock-2011}
Schollw\"ock,~U. The density-matrix renormalization group in the age of matrix
  product states. \emph{Annals of Physics} \textbf{2011}, \emph{326}, 96 --
  192, January 2011 Special Issue\relax
\mciteBstWouldAddEndPuncttrue
\mciteSetBstMidEndSepPunct{\mcitedefaultmidpunct}
{\mcitedefaultendpunct}{\mcitedefaultseppunct}\relax
\EndOfBibitem
\bibitem[{\relax Sz}alay \latin{et~al.}(2015){\relax Sz}alay, Pfeffer, Murg,
  Barcza, Verstraete, Schneider, and Legeza]{Szalay-2015a}
{\relax Sz}alay,~{\relax Sz}.; Pfeffer,~M.; Murg,~V.; Barcza,~G.;
  Verstraete,~F.; Schneider,~R.; Legeza,~{\"O}. Tensor product methods and
  entanglement optimization for ab initio quantum chemistry. \emph{Int. J.
  Quantum Chem.} \textbf{2015}, \emph{115}, 1342--1391\relax
\mciteBstWouldAddEndPuncttrue
\mciteSetBstMidEndSepPunct{\mcitedefaultmidpunct}
{\mcitedefaultendpunct}{\mcitedefaultseppunct}\relax
\EndOfBibitem
\bibitem[Verstraete \latin{et~al.}(2008)Verstraete, Murg, and
  Cirac]{Verstraete-2008}
Verstraete,~F.; Murg,~V.; Cirac,~J. Matrix product states, projected entangled
  pair states, and variational renormalization group methods for quantum spin
  systems. \emph{Advances in Physics} \textbf{2008}, \emph{57}, 143--224\relax
\mciteBstWouldAddEndPuncttrue
\mciteSetBstMidEndSepPunct{\mcitedefaultmidpunct}
{\mcitedefaultendpunct}{\mcitedefaultseppunct}\relax
\EndOfBibitem
\bibitem[Or{\'u}s(2019)]{Orus-2019}
Or{\'u}s,~R. Tensor networks for complex quantum systems. \emph{Nature Reviews
  Physics} \textbf{2019}, \emph{1}, 538--550\relax
\mciteBstWouldAddEndPuncttrue
\mciteSetBstMidEndSepPunct{\mcitedefaultmidpunct}
{\mcitedefaultendpunct}{\mcitedefaultseppunct}\relax
\EndOfBibitem
\bibitem[Chan \latin{et~al.}(2016)Chan, Keselman, Nakatani, Li, and
  White]{Chan-2020}
Chan,~G. K.-L.; Keselman,~A.; Nakatani,~N.; Li,~Z.; White,~S.~R. Matrix Product
  Operators, Matrix Product States, and ab initio Density Matrix
  Renormalization Group algorithms. 2016\relax
\mciteBstWouldAddEndPuncttrue
\mciteSetBstMidEndSepPunct{\mcitedefaultmidpunct}
{\mcitedefaultendpunct}{\mcitedefaultseppunct}\relax
\EndOfBibitem
\bibitem[Baiardi and Reiher(2020)Baiardi, and Reiher]{Baiardi-2020}
Baiardi,~A.; Reiher,~M. The density matrix renormalization group in chemistry
  and molecular physics: Recent developments and new challenges. \emph{The
  Journal of Chemical Physics} \textbf{2020}, \emph{152}, 040903\relax
\mciteBstWouldAddEndPuncttrue
\mciteSetBstMidEndSepPunct{\mcitedefaultmidpunct}
{\mcitedefaultendpunct}{\mcitedefaultseppunct}\relax
\EndOfBibitem
\bibitem[Cirac \latin{et~al.}(2021)Cirac, P\'erez-Garc\'{\i}a, Schuch, and
  Verstraete]{Cirac-2021}
Cirac,~J.~I.; P\'erez-Garc\'{\i}a,~D.; Schuch,~N.; Verstraete,~F. Matrix
  product states and projected entangled pair states: Concepts, symmetries,
  theorems. \emph{Rev. Mod. Phys.} \textbf{2021}, \emph{93}, 045003\relax
\mciteBstWouldAddEndPuncttrue
\mciteSetBstMidEndSepPunct{\mcitedefaultmidpunct}
{\mcitedefaultendpunct}{\mcitedefaultseppunct}\relax
\EndOfBibitem
\bibitem[Verstraete \latin{et~al.}(2023)Verstraete, Nishino, Schollw{\"o}ck,
  Ba{\~n}uls, Chan, and Stoudenmire]{Verstraete-2023}
Verstraete,~F.; Nishino,~T.; Schollw{\"o}ck,~U.; Ba{\~n}uls,~M.~C.;
  Chan,~G.~K.; Stoudenmire,~M.~E. Density matrix renormalization group, 30
  years on. \emph{Nature Reviews Physics} \textbf{2023}, 1--4\relax
\mciteBstWouldAddEndPuncttrue
\mciteSetBstMidEndSepPunct{\mcitedefaultmidpunct}
{\mcitedefaultendpunct}{\mcitedefaultseppunct}\relax
\EndOfBibitem
\bibitem[Menczer \latin{et~al.}(2024)Menczer, van Damme, Rask, Huntington,
  Hammond, Xantheas, Ganahl, and Örs Legeza]{Menczer-2024c}
Menczer,~A.; van Damme,~M.; Rask,~A.; Huntington,~L.; Hammond,~J.;
  Xantheas,~S.~S.; Ganahl,~M.; Örs Legeza, Parallel implementation of the
  Density Matrix Renormalization Group method achieving a quarter petaFLOPS
  performance on a single DGX-H100 GPU node. 2024;
  \url{https://arxiv.org/abs/2407.07411}\relax
\mciteBstWouldAddEndPuncttrue
\mciteSetBstMidEndSepPunct{\mcitedefaultmidpunct}
{\mcitedefaultendpunct}{\mcitedefaultseppunct}\relax
\EndOfBibitem
\bibitem[Swart and Gruden(2016)Swart, and Gruden]{Swart-2016}
Swart,~M.; Gruden,~M. Spinning around in Transition-Metal Chemistry.
  \emph{Accounts of Chemical Research} \textbf{2016}, \emph{49}, 2690--2697,
  PMID: 27993008\relax
\mciteBstWouldAddEndPuncttrue
\mciteSetBstMidEndSepPunct{\mcitedefaultmidpunct}
{\mcitedefaultendpunct}{\mcitedefaultseppunct}\relax
\EndOfBibitem
\bibitem[Khedkar and Roemelt(2021)Khedkar, and Roemelt]{Khedkar-2021}
Khedkar,~A.; Roemelt,~M. Modern multireference methods and their application in
  transition metal chemistry. \emph{Phys. Chem. Chem. Phys.} \textbf{2021},
  \emph{23}, 17097--17112\relax
\mciteBstWouldAddEndPuncttrue
\mciteSetBstMidEndSepPunct{\mcitedefaultmidpunct}
{\mcitedefaultendpunct}{\mcitedefaultseppunct}\relax
\EndOfBibitem
\bibitem[Feldt and Phung(2022)Feldt, and Phung]{Feldt-2022}
Feldt,~M.; Phung,~Q.~M. Ab Initio Methods in First-Row Transition Metal
  Chemistry. \emph{European Journal of Inorganic Chemistry} \textbf{2022},
  \emph{2022}, e202200014\relax
\mciteBstWouldAddEndPuncttrue
\mciteSetBstMidEndSepPunct{\mcitedefaultmidpunct}
{\mcitedefaultendpunct}{\mcitedefaultseppunct}\relax
\EndOfBibitem
\bibitem[Pantazis \latin{et~al.}(2009)Pantazis, Orio, Petrenko, Zein, Bill,
  Lubitz, Messinger, and Neese]{Pantazis-2009}
Pantazis,~D.; Orio,~M.; Petrenko,~T.; Zein,~S.; Bill,~E.; Lubitz,~W.;
  Messinger,~J.; Neese,~F. A New Quantum Chemical Approach to the Magnetic
  Properties of Oligonuclear Transition-Metal Complexes: Application to a Model
  for the Tetranuclear Manganese Cluster of Photosystem II. \emph{Chemistry –
  A European Journal} \textbf{2009}, \emph{15}, 5108--5123\relax
\mciteBstWouldAddEndPuncttrue
\mciteSetBstMidEndSepPunct{\mcitedefaultmidpunct}
{\mcitedefaultendpunct}{\mcitedefaultseppunct}\relax
\EndOfBibitem
\bibitem[Sharma \latin{et~al.}(2014)Sharma, Sivalingam, Neese, and
  Chan]{Sharma-2014}
Sharma,~S.; Sivalingam,~K.; Neese,~F.; Chan,~G. K.-L. Low-energy spectrum of
  iron--sulfur clusters directly from many-particle quantum mechanics.
  \emph{Nature Chemistry} \textbf{2014}, \emph{6}, 927--933\relax
\mciteBstWouldAddEndPuncttrue
\mciteSetBstMidEndSepPunct{\mcitedefaultmidpunct}
{\mcitedefaultendpunct}{\mcitedefaultseppunct}\relax
\EndOfBibitem
\bibitem[Vinyard and Brudvig(2017)Vinyard, and Brudvig]{Vinyard-2017}
Vinyard,~D.~J.; Brudvig,~G.~W. Progress Toward a Molecular Mechanism of Water
  Oxidation in Photosystem II. \emph{Annual Review of Physical Chemistry}
  \textbf{2017}, \emph{68}, 101--116, PMID: 28226223\relax
\mciteBstWouldAddEndPuncttrue
\mciteSetBstMidEndSepPunct{\mcitedefaultmidpunct}
{\mcitedefaultendpunct}{\mcitedefaultseppunct}\relax
\EndOfBibitem
\bibitem[Lubitz \latin{et~al.}(2019)Lubitz, Chrysina, and Cox]{Lubitz-2019}
Lubitz,~W.; Chrysina,~M.; Cox,~N. Water oxidation in photosystem II.
  \emph{Photosynthesis Research} \textbf{2019}, \emph{142}, 105--125\relax
\mciteBstWouldAddEndPuncttrue
\mciteSetBstMidEndSepPunct{\mcitedefaultmidpunct}
{\mcitedefaultendpunct}{\mcitedefaultseppunct}\relax
\EndOfBibitem
\bibitem[Kerridge(2015)]{Kerridge-2015}
Kerridge,~A. \emph{Computational Methods in Lanthanide and Actinide Chemistry};
  John Wiley \& Sons, Ltd, 2015; Chapter 5, pp 121--146\relax
\mciteBstWouldAddEndPuncttrue
\mciteSetBstMidEndSepPunct{\mcitedefaultmidpunct}
{\mcitedefaultendpunct}{\mcitedefaultseppunct}\relax
\EndOfBibitem
\bibitem[Spivak \latin{et~al.}(2017)Spivak, Vogiatzis, Cramer, Graaf, and
  Gagliardi]{Spivak-2017}
Spivak,~M.; Vogiatzis,~K.~D.; Cramer,~C.~J.; Graaf,~C.~d.; Gagliardi,~L.
  Quantum Chemical Characterization of Single Molecule Magnets Based on
  Uranium. \emph{The Journal of Physical Chemistry A} \textbf{2017},
  \emph{121}, 1726--1733, PMID: 28128563\relax
\mciteBstWouldAddEndPuncttrue
\mciteSetBstMidEndSepPunct{\mcitedefaultmidpunct}
{\mcitedefaultendpunct}{\mcitedefaultseppunct}\relax
\EndOfBibitem
\bibitem[Gaggioli and Gagliardi(2018)Gaggioli, and Gagliardi]{Gaggioli-2018}
Gaggioli,~C.~A.; Gagliardi,~L. Theoretical Investigation of Plutonium-Based
  Single-Molecule Magnets. \emph{Inorganic Chemistry} \textbf{2018}, \emph{57},
  8098--8105\relax
\mciteBstWouldAddEndPuncttrue
\mciteSetBstMidEndSepPunct{\mcitedefaultmidpunct}
{\mcitedefaultendpunct}{\mcitedefaultseppunct}\relax
\EndOfBibitem
\bibitem[Saue(2011)]{Trond-2011}
Saue,~T. Relativistic Hamiltonians for Chemistry: A Primer. \emph{ChemPhysChem}
  \textbf{2011}, \emph{12}, 3077--3094\relax
\mciteBstWouldAddEndPuncttrue
\mciteSetBstMidEndSepPunct{\mcitedefaultmidpunct}
{\mcitedefaultendpunct}{\mcitedefaultseppunct}\relax
\EndOfBibitem
\bibitem[Pyykk{\"o}(2012)]{Pyykk-2012}
Pyykk{\"o},~P. Relativistic effects in chemistry: more common than you thought.
  \emph{Annual review of physical chemistry} \textbf{2012}, \emph{63},
  45--64\relax
\mciteBstWouldAddEndPuncttrue
\mciteSetBstMidEndSepPunct{\mcitedefaultmidpunct}
{\mcitedefaultendpunct}{\mcitedefaultseppunct}\relax
\EndOfBibitem
\bibitem[Reiher and Wolf(2014)Reiher, and Wolf]{Reiher-2014a}
Reiher,~M.; Wolf,~A. \emph{Relativistic Quantum Chemistry}; John Wiley \& Sons,
  Ltd, 2014; Chapter 14, pp 527--566\relax
\mciteBstWouldAddEndPuncttrue
\mciteSetBstMidEndSepPunct{\mcitedefaultmidpunct}
{\mcitedefaultendpunct}{\mcitedefaultseppunct}\relax
\EndOfBibitem
\bibitem[Tecmer \latin{et~al.}(2016)Tecmer, Boguslawski, and
  K{{e}}dziera]{Tecmer-2016}
Tecmer,~P.; Boguslawski,~K.; K{{e}}dziera,~D. In \emph{Handbook of
  Computational Chemistry}; Leszczynski,~J., Ed.; Springer Netherlands:
  Dordrecht, 2016; pp 1--43\relax
\mciteBstWouldAddEndPuncttrue
\mciteSetBstMidEndSepPunct{\mcitedefaultmidpunct}
{\mcitedefaultendpunct}{\mcitedefaultseppunct}\relax
\EndOfBibitem
\bibitem[Liu(2020)]{Wenjian-2020}
Liu,~W. Essentials of relativistic quantum chemistry. \emph{The Journal of
  Chemical Physics} \textbf{2020}, \emph{152}, 180901\relax
\mciteBstWouldAddEndPuncttrue
\mciteSetBstMidEndSepPunct{\mcitedefaultmidpunct}
{\mcitedefaultendpunct}{\mcitedefaultseppunct}\relax
\EndOfBibitem
\bibitem[Zheng \latin{et~al.}(2017)Zheng, Chung, Corboz, Ehlers, Qin, Noack,
  Shi, White, Zhang, and Chan]{Zhang-2017}
Zheng,~B.-X.; Chung,~C.-M.; Corboz,~P.; Ehlers,~G.; Qin,~M.-P.; Noack,~R.~M.;
  Shi,~H.; White,~S.~R.; Zhang,~S.; Chan,~G. K.-L. Stripe order in the
  underdoped region of the two-dimensional Hubbard model. \emph{Science}
  \textbf{2017}, \emph{358}, 1155--1160\relax
\mciteBstWouldAddEndPuncttrue
\mciteSetBstMidEndSepPunct{\mcitedefaultmidpunct}
{\mcitedefaultendpunct}{\mcitedefaultseppunct}\relax
\EndOfBibitem
\bibitem[Menczer and Örs Legeza(2023)Menczer, and Örs Legeza]{Menczer-2023a}
Menczer,~A.; Örs Legeza, Massively Parallel Tensor Network State Algorithms on
  Hybrid CPU-GPU Based Architectures. \emph{arXiv:2305.05581} \textbf{2023},
  \relax
\mciteBstWouldAddEndPunctfalse
\mciteSetBstMidEndSepPunct{\mcitedefaultmidpunct}
{}{\mcitedefaultseppunct}\relax
\EndOfBibitem
\bibitem[Menczer and Legeza(2023)Menczer, and Legeza]{Menczer-2023b}
Menczer,~A.; Legeza,~{\"O}. Boosting the effective performance of massively
  parallel tensor network state algorithms on hybrid CPU-GPU based
  architectures via non-Abelian symmetries. 2023;
  \url{https://arxiv.org/abs/2309.16724}\relax
\mciteBstWouldAddEndPuncttrue
\mciteSetBstMidEndSepPunct{\mcitedefaultmidpunct}
{\mcitedefaultendpunct}{\mcitedefaultseppunct}\relax
\EndOfBibitem
\bibitem[Menczer \latin{et~al.}(2024)Menczer, Kap\'as, Werner, and
  Legeza]{Menczer-2024}
Menczer,~A.; Kap\'as,~K.; Werner,~M.~A.; Legeza,~{\"O}. Two-dimensional quantum
  lattice models via mode optimized hybrid CPU-GPU density matrix
  renormalization group method. \emph{Phys. Rev. B} \textbf{2024}, \emph{109},
  195148\relax
\mciteBstWouldAddEndPuncttrue
\mciteSetBstMidEndSepPunct{\mcitedefaultmidpunct}
{\mcitedefaultendpunct}{\mcitedefaultseppunct}\relax
\EndOfBibitem
\bibitem[Xiang \latin{et~al.}(2023)Xiang, Jia, Fang, and Li]{Xiang-2024}
Xiang,~C.; Jia,~W.; Fang,~W.-H.; Li,~Z. A distributed multi-GPU ab initio
  density matrix renormalization group algorithm with applications to the
  P-cluster of nitrogenase. 2023\relax
\mciteBstWouldAddEndPuncttrue
\mciteSetBstMidEndSepPunct{\mcitedefaultmidpunct}
{\mcitedefaultendpunct}{\mcitedefaultseppunct}\relax
\EndOfBibitem
\bibitem[NVIDIA(2023)]{nvidia-a100}
NVIDIA, Ampere (A100).
  \emph{\url{https://images.nvidia.com/aem-dam/en-zz/Solutions/data-center/nvidia-ampere-architecture-whitepaper.pdf}}
  \textbf{2023}, \relax
\mciteBstWouldAddEndPunctfalse
\mciteSetBstMidEndSepPunct{\mcitedefaultmidpunct}
{}{\mcitedefaultseppunct}\relax
\EndOfBibitem
\bibitem[NVIDIA(2023)]{nvidia-dgx-h100}
NVIDIA, NVIDIA DGX H100 TENSOR CORE GPU.
  \emph{\url{https://resources.nvidia.com/en-us-dgx-systems/ai-enterprise-dgx}}
  \textbf{2023}, \relax
\mciteBstWouldAddEndPunctfalse
\mciteSetBstMidEndSepPunct{\mcitedefaultmidpunct}
{}{\mcitedefaultseppunct}\relax
\EndOfBibitem
\bibitem[NVIDIA()]{gh200}
NVIDIA, {NVIDIA DGX GH200}.
  https://resources.nvidia.com/en-us-dgx-systems/nvidia-dgx-gh200-datasheet-web-us,
  \url{https://resources.nvidia.com/en-us-dgx-systems/nvidia-dgx-gh200-datasheet-web-us}\relax
\mciteBstWouldAddEndPuncttrue
\mciteSetBstMidEndSepPunct{\mcitedefaultmidpunct}
{\mcitedefaultendpunct}{\mcitedefaultseppunct}\relax
\EndOfBibitem
\bibitem[AMD()]{mi300}
AMD, {AMD INSTINCT MI300A APU}.
  https://www.amd.com/content/dam/amd/en/documents/instinct-tech-docs/data-sheets/amd-instinct-mi300a-data-sheet.pdf,
  \url{https://www.amd.com/content/dam/amd/en/documents/instinct-tech-docs/data-sheets/amd-instinct-mi300a-data-sheet.pdf}\relax
\mciteBstWouldAddEndPuncttrue
\mciteSetBstMidEndSepPunct{\mcitedefaultmidpunct}
{\mcitedefaultendpunct}{\mcitedefaultseppunct}\relax
\EndOfBibitem
\bibitem[Murg \latin{et~al.}(2010)Murg, Verstraete, Legeza, and
  Noack]{Murg-2010a}
Murg,~V.; Verstraete,~F.; Legeza,~{\"O}.; Noack,~R.~M. Simulating strongly
  correlated quantum systems with tree tensor networks. \emph{Phys. Rev. B}
  \textbf{2010}, \emph{82}, 205105\relax
\mciteBstWouldAddEndPuncttrue
\mciteSetBstMidEndSepPunct{\mcitedefaultmidpunct}
{\mcitedefaultendpunct}{\mcitedefaultseppunct}\relax
\EndOfBibitem
\bibitem[Nakatani and Chan(2013)Nakatani, and Chan]{Nakatani-2013}
Nakatani,~N.; Chan,~G. K.-L. Efficient tree tensor network states (TTNS) for
  quantum chemistry: Generalizations of the density matrix renormalization
  group algorithm. \emph{The Journal of Chemical Physics} \textbf{2013},
  \emph{138}\relax
\mciteBstWouldAddEndPuncttrue
\mciteSetBstMidEndSepPunct{\mcitedefaultmidpunct}
{\mcitedefaultendpunct}{\mcitedefaultseppunct}\relax
\EndOfBibitem
\bibitem[Murg \latin{et~al.}(2015)Murg, Verstraete, Schneider, Nagy, and
  Legeza]{Murg-2014}
Murg,~V.; Verstraete,~F.; Schneider,~R.; Nagy,~P.~R.; Legeza,~O. Tree Tensor
  Network State with Variable Tensor Order: An Efficient Multireference Method
  for Strongly Correlated Systems. 2015;
  \url{https://arxiv.org/abs/1403.0981}\relax
\mciteBstWouldAddEndPuncttrue
\mciteSetBstMidEndSepPunct{\mcitedefaultmidpunct}
{\mcitedefaultendpunct}{\mcitedefaultseppunct}\relax
\EndOfBibitem
\bibitem[Gunst \latin{et~al.}(2018)Gunst, Verstraete, Wouters, Legeza, and
  Van~Neck]{Gunst-2018}
Gunst,~K.; Verstraete,~F.; Wouters,~S.; Legeza,~{\"O}.; Van~Neck,~D. T3NS:
  Three-Legged Tree Tensor Network States. \emph{Journal of Chemical Theory and
  Computation} \textbf{2018}, \emph{14}, 2026--2033, PMID: 29481743\relax
\mciteBstWouldAddEndPuncttrue
\mciteSetBstMidEndSepPunct{\mcitedefaultmidpunct}
{\mcitedefaultendpunct}{\mcitedefaultseppunct}\relax
\EndOfBibitem
\bibitem[Gunst \latin{et~al.}(2019)Gunst, Verstraete, and Neck]{Gunst-2019}
Gunst,~K.; Verstraete,~F.; Neck,~D.~V. Three-Legged Tree Tensor Networks with
  SU(2) and Molecular Point Group Symmetry. \emph{Journal of Chemical Theory
  and Computation} \textbf{2019}, \emph{15}, 2996--3007\relax
\mciteBstWouldAddEndPuncttrue
\mciteSetBstMidEndSepPunct{\mcitedefaultmidpunct}
{\mcitedefaultendpunct}{\mcitedefaultseppunct}\relax
\EndOfBibitem
\bibitem[White(1992)]{White-1992b}
White,~S.~R. Density matrix formulation for quantum renormalization groups.
  \emph{Phys. Rev. Lett.} \textbf{1992}, \emph{69}, 2863--2866\relax
\mciteBstWouldAddEndPuncttrue
\mciteSetBstMidEndSepPunct{\mcitedefaultmidpunct}
{\mcitedefaultendpunct}{\mcitedefaultseppunct}\relax
\EndOfBibitem
\bibitem[Xiang(1996)]{Xiang-1996}
Xiang,~T. Density-matrix renormalization-group method in momentum space.
  \emph{Phys. Rev. B} \textbf{1996}, \emph{53}, R10445--R10448\relax
\mciteBstWouldAddEndPuncttrue
\mciteSetBstMidEndSepPunct{\mcitedefaultmidpunct}
{\mcitedefaultendpunct}{\mcitedefaultseppunct}\relax
\EndOfBibitem
\bibitem[White and Martin(1999)White, and Martin]{White-1999}
White,~S.~R.; Martin,~R.~L. Ab initio quantum chemistry using the density
  matrix renormalization group. \emph{The Journal of Chemical Physics}
  \textbf{1999}, \emph{110}, 4127--4130\relax
\mciteBstWouldAddEndPuncttrue
\mciteSetBstMidEndSepPunct{\mcitedefaultmidpunct}
{\mcitedefaultendpunct}{\mcitedefaultseppunct}\relax
\EndOfBibitem
\bibitem[Knecht \latin{et~al.}(2014)Knecht, Legeza, and Reiher]{Knecht-2014}
Knecht,~S.; Legeza,~{\"O}.; Reiher,~M. Communication: Four-component density
  matrix renormalization group. \emph{The Journal of Chemical Physics}
  \textbf{2014}, \emph{140}, 041101\relax
\mciteBstWouldAddEndPuncttrue
\mciteSetBstMidEndSepPunct{\mcitedefaultmidpunct}
{\mcitedefaultendpunct}{\mcitedefaultseppunct}\relax
\EndOfBibitem
\bibitem[Dukelsky and Pittel(2004)Dukelsky, and Pittel]{Dukelsky-2004}
Dukelsky,~J.; Pittel,~S. The density matrix renormalization group for finite
  {F}ermi systems. \emph{Reports on Progress in Physics} \textbf{2004},
  \emph{67}, 513\relax
\mciteBstWouldAddEndPuncttrue
\mciteSetBstMidEndSepPunct{\mcitedefaultmidpunct}
{\mcitedefaultendpunct}{\mcitedefaultseppunct}\relax
\EndOfBibitem
\bibitem[Legeza \latin{et~al.}(2015)Legeza, Veis, Poves, and
  Dukelsky]{Legeza-2015}
Legeza,~{\"O}.; Veis,~L.; Poves,~A.; Dukelsky,~J. Advanced density matrix
  renormalization group method for nuclear structure calculations. \emph{Phys.
  Rev. C} \textbf{2015}, \emph{92}, 051303\relax
\mciteBstWouldAddEndPuncttrue
\mciteSetBstMidEndSepPunct{\mcitedefaultmidpunct}
{\mcitedefaultendpunct}{\mcitedefaultseppunct}\relax
\EndOfBibitem
\bibitem[Legeza and Schilling(2018)Legeza, and Schilling]{Legeza-2018a}
Legeza,~{\"O}.; Schilling,~C. Role of the pair potential for the saturation of
  generalized Pauli constraints. \emph{Phys. Rev. A} \textbf{2018}, \emph{97},
  052105\relax
\mciteBstWouldAddEndPuncttrue
\mciteSetBstMidEndSepPunct{\mcitedefaultmidpunct}
{\mcitedefaultendpunct}{\mcitedefaultseppunct}\relax
\EndOfBibitem
\bibitem[Shapir \latin{et~al.}(2019)Shapir, Hamo, Pecker, Moca, Legeza, Zarand,
  and Ilani]{Shapir-2019}
Shapir,~I.; Hamo,~A.; Pecker,~S.; Moca,~C.~P.; Legeza,~{\"O}.; Zarand,~G.;
  Ilani,~S. Imaging the electronic Wigner crystal in one dimension.
  \emph{Science} \textbf{2019}, \emph{364}, 870--875\relax
\mciteBstWouldAddEndPuncttrue
\mciteSetBstMidEndSepPunct{\mcitedefaultmidpunct}
{\mcitedefaultendpunct}{\mcitedefaultseppunct}\relax
\EndOfBibitem
\bibitem[Barcza \latin{et~al.}(2021)Barcza, Ivády, Szilvási, Vörös, Veis,
  Ádám Gali, and Örs Legeza]{Barcza-2020}
Barcza,~G.; Ivády,~V.; Szilvási,~T.; Vörös,~M.; Veis,~L.; Ádám Gali,;
  Örs Legeza, DMRG on top of plane-wave Kohn-Sham orbitals: case study of
  defected boron nitride. \emph{J. Chem. Theory Comput.} \textbf{2021},
  \emph{17}, 1143--1154\relax
\mciteBstWouldAddEndPuncttrue
\mciteSetBstMidEndSepPunct{\mcitedefaultmidpunct}
{\mcitedefaultendpunct}{\mcitedefaultseppunct}\relax
\EndOfBibitem
\bibitem[Legeza \latin{et~al.}(2003)Legeza, R\"oder, and Hess]{Legeza-2003a}
Legeza,~{\"O}.; R\"oder,~J.; Hess,~B.~A. Controlling the accuracy of the
  density-matrix renormalization-group method: The dynamical block state
  selection approach. \emph{Phys. Rev. B} \textbf{2003}, \emph{67},
  125114\relax
\mciteBstWouldAddEndPuncttrue
\mciteSetBstMidEndSepPunct{\mcitedefaultmidpunct}
{\mcitedefaultendpunct}{\mcitedefaultseppunct}\relax
\EndOfBibitem
\bibitem[Krumnow \latin{et~al.}(2021)Krumnow, Veis, Eisert, and
  Legeza]{Krumnow-2021}
Krumnow,~C.; Veis,~L.; Eisert,~J.; Legeza,~{\"O}. Effective dimension reduction
  with mode transformations: Simulating two-dimensional fermionic condensed
  matter systems with matrix-product states. \emph{Phys. Rev. B} \textbf{2021},
  \emph{104}, 075137\relax
\mciteBstWouldAddEndPuncttrue
\mciteSetBstMidEndSepPunct{\mcitedefaultmidpunct}
{\mcitedefaultendpunct}{\mcitedefaultseppunct}\relax
\EndOfBibitem
\bibitem[Legeza and F\'ath(1996)Legeza, and F\'ath]{Legeza-1996}
Legeza,~{\"O}.; F\'ath,~G. Accuracy of the density-matrix renormalization-group
  method. \emph{Phys. Rev. B} \textbf{1996}, \emph{53}, 14349--14358\relax
\mciteBstWouldAddEndPuncttrue
\mciteSetBstMidEndSepPunct{\mcitedefaultmidpunct}
{\mcitedefaultendpunct}{\mcitedefaultseppunct}\relax
\EndOfBibitem
\bibitem[Wolf \latin{et~al.}(2008)Wolf, Verstraete, Hastings, and
  Cirac]{Wolf-2008}
Wolf,~M.~M.; Verstraete,~F.; Hastings,~M.~B.; Cirac,~J.~I. Area Laws in Quantum
  Systems: Mutual Information and Correlations. \emph{Physical Review Letters}
  \textbf{2008}, \emph{100}\relax
\mciteBstWouldAddEndPuncttrue
\mciteSetBstMidEndSepPunct{\mcitedefaultmidpunct}
{\mcitedefaultendpunct}{\mcitedefaultseppunct}\relax
\EndOfBibitem
\bibitem[Eisert \latin{et~al.}(2010)Eisert, Cramer, and Plenio]{Eisert-2010}
Eisert,~J.; Cramer,~M.; Plenio,~M.~B. Colloquium: Area laws for the
  entanglement entropy. \emph{Rev. Mod. Phys.} \textbf{2010}, \emph{82},
  277--306\relax
\mciteBstWouldAddEndPuncttrue
\mciteSetBstMidEndSepPunct{\mcitedefaultmidpunct}
{\mcitedefaultendpunct}{\mcitedefaultseppunct}\relax
\EndOfBibitem
\bibitem[Noack(2005)]{Noack-2005}
Noack,~R.~M. Diagonalization- and Numerical Renormalization-Group-Based Methods
  for Interacting Quantum Systems. {AIP} Conference Proceedings. 2005\relax
\mciteBstWouldAddEndPuncttrue
\mciteSetBstMidEndSepPunct{\mcitedefaultmidpunct}
{\mcitedefaultendpunct}{\mcitedefaultseppunct}\relax
\EndOfBibitem
\bibitem[Chan \latin{et~al.}(2008)Chan, Dorando, Ghosh, Hachmann, Neuscamman,
  Wang, and Yanai]{Chan-2008}
Chan,~G. K.-L.; Dorando,~J.~J.; Ghosh,~D.; Hachmann,~J.; Neuscamman,~E.;
  Wang,~H.; Yanai,~T. In \emph{Frontiers in Quantum Systems in Chemistry and
  Physics}; Wilson,~S., Grout,~P.~J., Maruani,~J., Delgado-Barrio,~G.,
  Piecuch,~P., Eds.; Progress in Theoretical Chemistry and Physics; Springer:
  Netherlands, 2008; Vol.~18\relax
\mciteBstWouldAddEndPuncttrue
\mciteSetBstMidEndSepPunct{\mcitedefaultmidpunct}
{\mcitedefaultendpunct}{\mcitedefaultseppunct}\relax
\EndOfBibitem
\bibitem[Or\'us(2014)]{Orus-2014}
Or\'us,~R. A practical introduction to tensor networks: Matrix product states
  and projected entangled pair states. \emph{Annals of Physics} \textbf{2014},
  \emph{349}, 117 -- 158\relax
\mciteBstWouldAddEndPuncttrue
\mciteSetBstMidEndSepPunct{\mcitedefaultmidpunct}
{\mcitedefaultendpunct}{\mcitedefaultseppunct}\relax
\EndOfBibitem
\bibitem[MAT()]{MATLAB}
MATLAB multi-paradigm programming language.
  \url{https://www.mathworks.com/}\relax
\mciteBstWouldAddEndPuncttrue
\mciteSetBstMidEndSepPunct{\mcitedefaultmidpunct}
{\mcitedefaultendpunct}{\mcitedefaultseppunct}\relax
\EndOfBibitem
\bibitem[boo()]{boostlib}
Boost C++ Libraries. \url{https://www.boost.org/}\relax
\mciteBstWouldAddEndPuncttrue
\mciteSetBstMidEndSepPunct{\mcitedefaultmidpunct}
{\mcitedefaultendpunct}{\mcitedefaultseppunct}\relax
\EndOfBibitem
\bibitem[foo()]{footnote-fetching}
The execution of the algorithm on the compute node is halted until the save IO
  operation by the slave process is completed.\relax
\mciteBstWouldAddEndPunctfalse
\mciteSetBstMidEndSepPunct{\mcitedefaultmidpunct}
{}{\mcitedefaultseppunct}\relax
\EndOfBibitem
\bibitem[Menczer and Legeza(2023)Menczer, and Legeza]{Menczer-2023c}
Menczer,~A.; Legeza,~{\"O}. Petaflops Density Matrix Renormalization Group
  Method, unpublished. 2023\relax
\mciteBstWouldAddEndPuncttrue
\mciteSetBstMidEndSepPunct{\mcitedefaultmidpunct}
{\mcitedefaultendpunct}{\mcitedefaultseppunct}\relax
\EndOfBibitem
\bibitem[Reiher \latin{et~al.}(2017)Reiher, Wiebe, Svore, Wecker, and
  Troyer]{Reiher-2017}
Reiher,~M.; Wiebe,~N.; Svore,~K.~M.; Wecker,~D.; Troyer,~M. Elucidating
  reaction mechanisms on quantum computers. \emph{Proceedings of the National
  Academy of Sciences} \textbf{2017}, \emph{114}, 7555--7560\relax
\mciteBstWouldAddEndPuncttrue
\mciteSetBstMidEndSepPunct{\mcitedefaultmidpunct}
{\mcitedefaultendpunct}{\mcitedefaultseppunct}\relax
\EndOfBibitem
\bibitem[Li \latin{et~al.}(2019)Li, Li, Dattani, Umrigar, and Chan]{Li-2019}
Li,~Z.; Li,~J.; Dattani,~N.~S.; Umrigar,~C.~J.; Chan,~G. K.-L. The electronic
  complexity of the ground-state of the FeMo cofactor of nitrogenase as
  relevant to quantum simulations. \emph{The Journal of Chemical Physics}
  \textbf{2019}, \emph{150}, 024302\relax
\mciteBstWouldAddEndPuncttrue
\mciteSetBstMidEndSepPunct{\mcitedefaultmidpunct}
{\mcitedefaultendpunct}{\mcitedefaultseppunct}\relax
\EndOfBibitem
\bibitem[Guther \latin{et~al.}(2020)Guther, Anderson, Blunt, Bogdanov, Cleland,
  Dattani, Dobrautz, Ghanem, Jeszenszki, Liebermann, Manni, Lozovoi, Luo, Ma,
  Merz, Overy, Rampp, Samanta, Schwarz, Shepherd, Smart, Vitale, Weser, Booth,
  and Alavi]{Kai-2020}
Guther,~K. \latin{et~al.}  NECI: N-Electron Configuration Interaction with an
  emphasis on state-of-the-art stochastic methods. \emph{The Journal of
  Chemical Physics} \textbf{2020}, \emph{153}, 034107\relax
\mciteBstWouldAddEndPuncttrue
\mciteSetBstMidEndSepPunct{\mcitedefaultmidpunct}
{\mcitedefaultendpunct}{\mcitedefaultseppunct}\relax
\EndOfBibitem
\bibitem[Brabec \latin{et~al.}(2021)Brabec, Brandejs, Kowalski, Xantheas,
  Legeza, and Veis]{Brabec-2021}
Brabec,~J.; Brandejs,~J.; Kowalski,~K.; Xantheas,~S.; Legeza,~{\"O}.; Veis,~L.
  Massively parallel quantum chemical density matrix renormalization group
  method. \emph{Journal of Computational Chemistry} \textbf{2021}, \emph{42},
  534--544\relax
\mciteBstWouldAddEndPuncttrue
\mciteSetBstMidEndSepPunct{\mcitedefaultmidpunct}
{\mcitedefaultendpunct}{\mcitedefaultseppunct}\relax
\EndOfBibitem
\bibitem[Hoffman \latin{et~al.}(2014)Hoffman, Lukoyanov, Yang, Dean, and
  Seefeldt]{Hoffman-2014}
Hoffman,~B.~M.; Lukoyanov,~D.; Yang,~Z.-Y.; Dean,~D.~R.; Seefeldt,~L.~C.
  Mechanism of Nitrogen Fixation by Nitrogenase: The Next Stage. \emph{Chemical
  Reviews} \textbf{2014}, \emph{114}, 4041--4062, PMID: 24467365\relax
\mciteBstWouldAddEndPuncttrue
\mciteSetBstMidEndSepPunct{\mcitedefaultmidpunct}
{\mcitedefaultendpunct}{\mcitedefaultseppunct}\relax
\EndOfBibitem
\bibitem[hlr()]{hlrs}
HPE Hawk Hardware and Architecture.
  \url{https://kb.hlrs.de/platforms/index.php/HPE_Hawk_Hardware_and_Architecture#Interconnect}\relax
\mciteBstWouldAddEndPuncttrue
\mciteSetBstMidEndSepPunct{\mcitedefaultmidpunct}
{\mcitedefaultendpunct}{\mcitedefaultseppunct}\relax
\EndOfBibitem
\end{mcitethebibliography}


\end{document}